\documentclass[aps, prl, reprint]{revtex4-2}
\usepackage[british]{babel}
\usepackage{times}

\usepackage{amssymb}
\usepackage{mathtools}
\usepackage{physics} 
\usepackage{siunitx}

\usepackage[dvipsnames]{xcolor}
\definecolor{myblue}{RGB}{34,31,150}

\usepackage[breaklinks=true,colorlinks=true,linkcolor=myblue,urlcolor=myblue,citecolor=myblue]{hyperref}

\begin{document}

% Title, authors, institutions and date
\title{First-principles construction of symmetry-informed quantum metrologies}
\author{Jes\'{u}s Rubio}
\email{j.rubiojimenez@surrey.ac.uk}
\affiliation{School of Mathematics and Physics, University of Surrey, Guildford GU2 7XH, United Kingdom}
\date{\today}
   
% Abstract
\begin{abstract}

Combining quantum and Bayesian principles leads to optimality in metrology, but the optimisation equations involved are often hard to solve. 
This work mitigates this problem with a novel class of measurement strategies for quantities isomorphic to location parameters, which are shown to admit a closed-form optimisation. 
The resulting framework admits any parameter range, prior information, or state, and the associated estimators apply to finite samples. 
As an example, the metrology of relative weights is formulated from first principles and shown to require hyperbolic errors.  
The primary advantage of this approach lies in its simplifying power: it reduces the search for good strategies to identifying which symmetry leaves a state of maximum ignorance invariant.
This will facilitate the application of quantum metrology to fundamental physics, where symmetries play a key role. 

\end{abstract}

\maketitle

% Body
Quantum metrology is often thought of as inseparable from phase estimation \cite{demkowicz2015quantum}. 
This has led to numerous insights in the foundations of physics, including a $\pi$-corrected Heisenberg limit \cite{gorecki2020pi}, clarification of the role of entanglement in quantum-enhanced measurements \cite{sahota2015quantum, proctor2017multiparameter, braun2018quantum, bringewatt2021protocols}, the construction of phase observables \cite{luis1996optimum}, and the enhancement of gravitational wave detection via squeezed light \cite{caves1981quantum-mechanical, lough2021first, heinze2022quantum}. 
Yet, modern quantum technologies \cite{degen2017quantum, vinjanampathy2016quantum, belenchia2021quantum} are inspiring metrology problems that transcend phase estimation.

Two examples are quantum thermometry \cite{stace2010quantum, mehboudi2019thermometry, rubio2021global, mehboudi2021fundamental} and rate estimation in dissipative processes \cite{vidrighin2014joint, branford2019quantum}. 
Since temperature and rate set energy and time scales \cite{jaynes1968prior, prosper1993temperature}, respectively, scale invariance becomes essential for consistent estimation, and this reveals a metrological framework for scales that is independent of phase estimation \cite{rubio2023quantum}.
This strongly suggests that different parameter types require different metrologies.

Far from a mere formality, this idea is proving crucial in the presence of finite information \cite{lumino2017experimental, rubio2018quantum, morelli2021bayesian, meyer2023quantum}.
The application of scale estimation to a thermometry experiment on $^{41}\mathrm{K}$ atoms confined in an optical tweezer at microkelvin temperatures \cite{glatthard2022optimal}, for instance, has demonstrated how individual measurements can be made substantially more informative by enforcing the correct invariance via Bayesian principles \cite{jaynes2003probability, linden2014bayesian}.

We have circular invariance for phases \cite{holevo2011probabilistic, chiribella2005optimal, demkowicz2011optimal, chesi2023protocol}, translation invariance for locations \cite{personick1971application}, and scale invariance for scales \cite{rubio2023quantum}. 
But not every parameter falls under such categories.
This is the case, e.g., of relative weights, that is, any $\eta \in (0, 1)$ quantifying the relative importance of any two objects as $\eta$ and $1-\eta$.
Examples include probability of success \cite{jaynes1968prior}, blend parameters in mixed states \cite{englert2013on}, photon loss in an interferometer \cite{crowley2014tradeoff}, and the Schmidt parameter characterising the class of two-qubit pure states and their entanglement \cite{genoni2008optimal}.

A simple, yet effective way of discovering new metrologies is to exploit the class of parameters that can be mapped into locations. 
If $\Theta$ is such a parameter, and its value is completely unknown, there will exist a function $f\hspace{-0.3em}: \theta \rightarrow f(\theta)$ such that our initial state of knowledge is invariant under transformations
\begin{equation}
  f(\theta) \mapsto f(\theta') = f(\theta) + c,
  \label{eq:basic-transformation}
\end{equation}
for arbitrary $c$ and where $\theta$ and $\theta'$ denote different but equally valid hypotheses about $\Theta$. 
Such hypotheses are related by a transformation $\theta' = g(\theta)$ that is determined by the physics at hand, and we say that $f$ maps $\Theta$ into a location because Eq.~\eqref{eq:basic-transformation} is a translation of $f(\theta)$ \cite{jaynes2003probability}.
Good measurement strategies can then be found by optimising the family of \emph{quadratic} errors
\begin{equation}
  \mathcal{D}_{f}[\tilde{\theta}_{\boldsymbol{y}}(x),\theta] = \lbrace f[\tilde{\theta}_{\boldsymbol{y}}(x)] - f(\theta)\rbrace^2
  \label{eq:quadratic}
\end{equation}
on average, where $x$ and $\boldsymbol{y} = (y_1, y_2, \dots) $ denote a measurement outcome and control parameters, respectively, and the map $\tilde{\theta}_{\boldsymbol{y}}\hspace{-0.25em}: x \rightarrow \tilde{\theta}_{\boldsymbol{y}}(x)$ processes $x$ into an estimate for $\Theta$.
Since the form of $f$ depends on the transformation $g$, and this varies for different parameter types, imposing the symmetry \eqref{eq:basic-transformation} leads to different metrologies in different scenarios.
For example, $f(z) = z$ transforms Eq.~\eqref{eq:quadratic} into the square error $[\tilde{\theta}_{\boldsymbol{y}}(x) - \theta]^2$ used for locations, while $f(z) = \log(z/z_0)$, with constant $z_0$, renders the logarithmic error $\mathrm{log}^2[\tilde{\theta}_{\boldsymbol{y}}(x)/ \theta]$ used for scales \cite{rubio2023quantum}. 
In general, Eqs.~\eqref{eq:basic-transformation} and \eqref{eq:quadratic} cover a vast collection of parameter types largely unexplored.

This Letter reports an optimal framework for the quantum metrology of such location-isomorphic parameters.
Owing to its symmetry-informed nature, this approach provides closed-form rules to calculate, for given state and function $f$, optimal estimators and probability-operator measurements (POMs).
Since they are Bayesian, these rules are global, i.e., valid for a hypothesis range $\theta \in [\theta_{\mathrm{min}}, \theta_{\mathrm{max}}]$ as wide or as narrow as required (including the local regime $\theta_{\mathrm{max}}/\theta_{\mathrm{min}} \sim 1$ \cite{demkowicz2015quantum}).
Moreover, the calculation of the associated minimum errors is exact, thus bypassing the hierarchies of error bounds typically employed in metrology \cite{tsang2012zivzakai, tsang2016quantum, demkowicz2020multiparameter, suzuki2023bayesian, chang2023global} (including the celebrated Cramér-Rao bounds that cannot always be applied \footnote{
Despite the extensive use of Cramér-Rao bounds in metrology \cite{helstrom1976quantum, demkowicz2015quantum}, their saturation requires either exponential family distributions or locally unbiased estimators in narrow parameter ranges. 
Even if we only approach these bounds, an asymptotically large number of measurements is generally needed. 
This contrasts with the universal applicability of Bayesian techniques, which are particularly useful in experimental configurations with broad prior ranges, finite samples, and distributions that do not belong to the exponential family \cite{lumino2017experimental, rubio2018quantum, morelli2021bayesian, meyer2023quantum}.
While Bayesian estimators are often `biased' as per the notion of unbiasedness of frequentist estimation \cite{teklu2009bayesian, brivio2010experimental}, this bears no relevance for Bayesian frameworks where estimator optimality is established by the direct minimisation of a suitable uncertainty quantifier.
Moreover, the bias of an estimator is defined with respect to the mean square error, while this work introduces other error types for which such a definition ceases to apply.
}).  
This generalises Personick's pioneering work \cite{personick1971application}---an early demonstration of the fruitful marriage between quantum detection and Bayesian inference \cite{helstrom1976quantum}. 
Personick's discovery of a quantum minimum square error has been refined \cite{helstrom1976quantum, macieszczak2014bayesian}, extended \cite{chabuda2016the, rubio2020bayesian, sidhu2020geometric, rubio2023quantum, suzuki2023bayesian}, and applied \cite{mashide2002optimal, wang2007quantum, ban2016bayes, mazurek2016sharp, sekatski2017, bernad2018optimal, rubio2018quantum, nichols2019designing, rexiti2019adversarial, branford2021average, fanizza2020beyond, lee2023quantum, zhou2023bayesianA, zhou2023bayesianB, bavaresco2023designing} over half a century. 
But, aside from scale estimation \cite{rubio2023quantum}, his reasoning has seemingly been restricted to location metrology.  

To demonstrate the power of this framework, a metrology of weights is derived from first principles. 
The symmetry \eqref{eq:basic-transformation} is shown to arise in this case from Möbius transformations, leading to a hyperbolic error and estimators based on the logistic function.
Their application to the estimation of a blend parameter in a mixed state reveals a maximum precision gain of $75 \%$ relative to the prior uncertainty. 
This illustrates the key advantage of symmetry-informed estimation: if $\Theta$ is location-isomorphic, finding the best estimator and POM amounts to identifying the symmetry \eqref{eq:basic-transformation} and performing a \emph{single} calculation of the optimal strategy using the rules reported here.

\emph{Elements of quantum metrology.}---We wish to estimate $\Theta$.
A finite hypothesis range $\theta \in [\theta_{\mathrm{min}},\theta_{\mathrm{max}}]$ is often available in practice, and other kinds of prior knowledge can be accounted for using principles such as maximum entropy \cite{jaynes2003probability, presse2013principles}. 
If, on the other hand, we start from minimal assumptions including the type of parameter $\Theta$ is---a scale, a weight, etc---and its general support, then we are \emph{maximally ignorant} about its value \cite{jaynes1968prior}.
A prior probability $p(\theta)$ is used to encode any available (or the absence of) initial information.

The hypothesis $\theta$ is next encoded in a state $\rho_{\boldsymbol{y}}(\theta)$, which is often characterised by some control parameters $\boldsymbol{y}$.
Examples include preparation and readout times in magnetic field sensing \cite{hayes2018making}, and expansion time in release-recapture thermometry \cite{glatthard2022optimal}.
A POM $M_{\boldsymbol{y}}(x)$ is performed on this state, and the outcome $x$ is used to update the information in $p(\theta)$. 
This procedure provides the desired estimate $\tilde{\theta}_{\boldsymbol{y}} \pm \Delta \tilde{\theta}_{\boldsymbol{y}}$, where $\Delta \tilde{\theta}_{\boldsymbol{y}}$ denotes an outcome-dependent error. 
 
A central problem in this context is finding estimators and POMs leading to the least error.
The next section provides an exact, analytical solution to this for quadratic errors [Eq.~\eqref{eq:quadratic}].

\emph{Optimal strategy for quadratic errors.}---We start by integrating Eq.~\eqref{eq:quadratic} weighted over $\theta$ and $x$ as
\begin{equation}
   \mathrm{Tr}\left\lbrace \int dx M_{\boldsymbol{y}}(x) W_f [\tilde{\theta}_{\boldsymbol{y}}(x)]\right\rbrace
   \coloneqq 
   \bar{\epsilon}_{\boldsymbol{y}, f, \mathrm{MQE}},
\label{eq:mqe}
\end{equation}
where 
\begin{equation}
  W_f [\tilde{\theta}_{\boldsymbol{y}}(x)] = \int d\theta\,p(\theta) \rho_{\boldsymbol{y}}(\theta)\lbrace f[\tilde{\theta}_{\boldsymbol{y}}(x)] - f(\theta)\rbrace^2.
\end{equation}
We average over the hypothesis $\theta$ because $\Theta$ is unknown; this makes the error globally valid, i.e., for any parameter range.
Similarly, we average over the outcome $x$ because the search for optimal POMs takes place prior to recording a specific measurement outcome.
Eq.~\eqref{eq:mqe} is a \emph{mean quadratic error}.

To find the optimal strategy minimising this error, it is useful to rewrite it as 
\begin{equation}
  \epsilon_{\boldsymbol{y},f,\mathrm{MQE}} = \zeta_f + \mathrm{Tr}(\rho_{\boldsymbol{y},f,0} \mathcal{A}_{\boldsymbol{y},f,2} - 2\rho_{\boldsymbol{y},f,1} \mathcal{A}_{\boldsymbol{y},f,1}),
\label{eq:mqe-simpler}
\end{equation}
where
\begin{subequations}
  \begin{equation}
    \zeta_f = \int d\theta\,p(\theta) f(\theta)^2,
  \end{equation}
  \begin{equation}
    \rho_{\boldsymbol{y},f,l} = \int d\theta\,p(\theta) \rho_{\boldsymbol{y}}(\theta) f(\theta)^l,
  \label{eq:state-aux}
  \end{equation}
  \begin{equation}
    \mathcal{A}_{\boldsymbol{y},f,l} = \int dx M_{\boldsymbol{y}}(x) f[\tilde{\theta}_{\boldsymbol{y}}(x)]^l.
  \label{eq:pom-aux}
  \end{equation}
\end{subequations}
By virtue of Jensen's inequality, $\mathcal{A}_{\boldsymbol{y},f,2} - \mathcal{A}_{\boldsymbol{y},f,1}^2 \geq 0$, Eq.~\eqref{eq:mqe-simpler} is lower bounded as 
\begin{equation}
  \epsilon_{\boldsymbol{y},f,\mathrm{MQE}} \geq \zeta_f + \mathrm{Tr}(\rho_{\boldsymbol{y},f,0} \mathcal{A}_{\boldsymbol{y},f,1}^2 - 2\rho_{\boldsymbol{y},f,1} \mathcal{A}_{\boldsymbol{y},f,1}).
  \label{eq:mqe-bound}
\end{equation}
But projective measurements---i.e., $M_{\boldsymbol{y}}(x)M_{\boldsymbol{y}}(x')\rightarrow \delta(x -x') M_{\boldsymbol{y}}(x')$---saturate Jensen's inequality.
Therefore, we can assume equality in Eq.~\eqref{eq:mqe-bound} and restrict the search to projective strategies without loss of optimality \cite{macieszczak2014bayesian}.
 
Using variational calculus, and following the formally analogous derivation in Refs~\cite{personick1971application,rubio2023quantum}, such an equality is found to achieve its minimum at
\begin{equation}
  \mathcal{A}_{\boldsymbol{y},f,1} = \mathcal{S}_{\boldsymbol{y},f},
\label{eq:min-cond}
\end{equation}
where $\mathcal{S}_{\boldsymbol{y},f}$ solves the Lyapunov equation
\begin{equation}
  \mathcal{S}_{\boldsymbol{y},f}\rho_{\boldsymbol{y},f,0} + \rho_{\boldsymbol{y},f,0}\mathcal{S}_{\boldsymbol{y},f} = 2\rho_{\boldsymbol{y},f,1}.
\label{eq:lyaponuv}
\end{equation}
Crucially, $\mathcal{S}_{\boldsymbol{y},f}$ contains all the information about the optimal strategy, as follows. 
Given the eigendecomposition
\begin{equation}
  \mathcal{S}_{\boldsymbol{y},f} = \int ds \mathcal{P}_{\boldsymbol{y},f}(s) s,
  \label{eq:eigendecomposition}
\end{equation}
where $\mathcal{P}_{\boldsymbol{y},f}(s)\mathcal{P}_{\boldsymbol{y},f}(s')\rightarrow \delta(s -s') \mathcal{P}_{\boldsymbol{y},f}(s')$, and recalling the definition in Eq.~\eqref{eq:pom-aux}, Eq.~\eqref{eq:min-cond} implies
\begin{subequations}\label{eq:optimal-strategy}
  \begin{equation}
    \tilde{\theta}_{\boldsymbol{y}}(x) \mapsto f^{-1}(s) \coloneqq \tilde{\vartheta}_{\boldsymbol{y},f}(s),
  \label{eq:op-est}
  \end{equation}
  \begin{equation}
    M_{\boldsymbol{y}}(x) \mapsto  \mathcal{P}_{\boldsymbol{y},f}(s) \coloneq \mathcal{M}_{\boldsymbol{y},f}(s).
  \label{eq:op-pom}
  \end{equation}
\end{subequations}
The optimal estimator is thus found by transforming the spectrum of $\mathcal{S}_{\boldsymbol{y},f}$ via the inverse $f$ map [Eq.~\eqref{eq:op-est}], while the optimal measurement consists in projecting onto the eigenspace of $\mathcal{S}_{\boldsymbol{y},f}$ [Eq.~\eqref{eq:op-pom}].

Inserting Eq.~\eqref{eq:min-cond} into Eq.~\eqref{eq:mqe-bound} further renders the associated minimum error  
\begin{equation}
  \bar{\epsilon}_{\boldsymbol{y},f,\mathrm{min}} = \bar{\epsilon}_{p,f} - \mathcal{G}_{\boldsymbol{y},f}
  \label{eq:min}
\end{equation}
as the difference between the initial uncertainty $\bar{\epsilon}_{p,f}$, given by the prior variance of $f$, and the average precision gain
\begin{equation}
  \mathcal{G}_{\boldsymbol{y},f} = \mathrm{Tr}(\rho_{\boldsymbol{y},f,0} \mathcal{S}_{\boldsymbol{y},f}^2) - \mathrm{Tr}(\rho_{\boldsymbol{y},f,0} \mathcal{S}_{\boldsymbol{y},f})^2.
\label{eq:gain}
\end{equation}
Eq.~\eqref{eq:min} is useful, in addition, to assess the relative performance of suboptimal---but perhaps more practical---strategies via the trivial uncertainty relation $\bar{\epsilon}_{\boldsymbol{y},f,\mathrm{MQE}} \geq \bar{\epsilon}_{\boldsymbol{y},f,\mathrm{min}}$.

Eqs.~\eqref{eq:lyaponuv}, \eqref{eq:optimal-strategy}, and \eqref{eq:min} are the main result of this work. 
They generalise Personick's framework \cite{personick1971application} (as well as scale estimation \cite{rubio2023quantum}) and provide the optimal quantum strategy for any location-isomorphic parameter.
This is next illustrated for weight parameters.

\emph{Weight estimation.}---
Consider a set with two generic elements, $e_0$ and $e_1$, carrying weights $\eta$ and $1 - \eta$, respectively.
Suppose $\eta$ is unknown.
To construct a quantum metrology for $\eta$, we first need a notion of maximum ignorance.

Let $\theta \in (0, 1)$ be a hypothesis about $\eta$.
If we ask how likely it is that one would choose $e_0$ over $e_1$, our probability for this is $p(e_0) = \theta$.
If a new piece of information $I$ is provided---$I$ denotes a proposition---$p(e_0)$ can be updated to $p(e_0|I)$ via Bayes's theorem. 
But $p(e_0|I) = \theta'$ can also be used as a hypothesis for $\eta$.
This induces a Möbius transformation
\begin{equation}
  \theta' = \frac{\gamma\theta}{1 - \theta + \gamma\theta} 
  \label{eq:mobius}
\end{equation}
between hypotheses, with $\gamma = p(I|e_0)/p(I|e_1)$. 
By rewriting it as $\theta'/(1-\theta') = \gamma \theta/(1-\theta)$, we see that it amounts to rescaling the `odds'.
But this rescaling does not inform the value of $\eta$.
We then say that our initial state of knowledge is invariant under odds transformations \eqref{eq:mobius}.
This motivates the formal constraint $p(\theta)d\theta = p(\theta')d\theta'$ on the prior probability, which renders the functional equation
\begin{equation}
(1 - \theta + \gamma\theta)^2 p(\theta) = \gamma p\left(\frac{\gamma\theta}{1 - \theta + \gamma\theta} \right).
\end{equation}
Its solution, $p(\theta) \propto 1/[\theta(1 - \theta)]$, is sometimes referred to as Haldane's prior \cite{haldane1932a}.
This derivation was suggested by Jaynes \cite{jaynes1968prior} for a probability of success, and it has here been extended to any weight parameter. 

Having found an ignorance prior for weights, an appropriate error can be derived. 
Setting $\varphi = c_1 \mathrm{artanh}(2\theta-1) + c_2$, with arbitrary $c_1$ and $c_2$, maps our weight estimation problem into that of finding a location with hypothesis $\varphi \in (-\infty, \infty)$.
Namely, $p(\theta) d\theta = p(\varphi) d\varphi$ implies $p(\theta) \propto 1/[\theta(1 - \theta)] \mapsto p(\varphi) \propto 1$, where $p(\varphi) \propto 1$ represents maximum ignorance about locations \cite{jaynes1968prior}. 
The deviation of $\tilde{\varphi}$ from $\varphi$ is quantified by the $k$ distance $\mathcal{D}_k(\tilde{\varphi},\varphi) = | \tilde{\varphi} - \varphi |^k$; consequently, 
\begin{equation}
  \mathcal{D}_k(\tilde{\varphi},\varphi) \mapsto 
  \mathcal{D}_k(\tilde{\theta},\theta) = 
  \Bigg\vert c_1 \mathrm{artanh}\left(\frac{\tilde{\theta} - \theta}{\tilde{\theta} + \theta - 2\tilde{\theta} \theta}\right) \Bigg\vert^k.
  \label{eq:hyperbolic-err}
\end{equation}
Eq.~\eqref{eq:hyperbolic-err} is symmetric, $\mathcal{D}_k(\tilde{\theta}, \theta) = \mathcal{D}_k(\theta, \tilde{\theta})$; invariant under Eq.~\eqref{eq:mobius}, $\mathcal{D}_k(\tilde{\theta}', \theta') = \mathcal{D}_k(\tilde{\theta}, \theta)$; it vanishes at $\tilde{\theta} = \theta$; and it grows (decreases) monotonically from (towards) that point. 
It is thus a \emph{bona fide} error for weights.
Once identified under minimal assumptions, it can be combined with prior probabilities other than Haldane's \cite{tuyl2008frank}.

We next apply symmetry-informed estimation. 
Let $c_1 = 2$ without loss of generality, and $k = 2$ for the error to be quadratic. 
This turns Eq.~\eqref{eq:mqe} into a \emph{mean hyperbolic error} with $f(z) = 2\,\mathrm{artanh}(2z-1)$. 
Applying this $f$ map to Eq.~\eqref{eq:mobius} reveals the translation symmetry $f(\theta') = f(\theta) + c$, with $c = \log(\gamma)$. 
Weight parameters are thus location-isomorphic.
This implies that
the best estimation strategy can be found by solving Eq.~\eqref{eq:lyaponuv}, for which Eq.~\eqref{eq:state-aux} takes the form
\begin{equation}
  \rho_{\boldsymbol{y},l} = 2^l \int d\theta\,p(\theta) \rho_{\boldsymbol{y}}(\theta)\,\mathrm{artanh}(2\theta - 1)^l.
\label{eq:hyperbolic-aux}
\end{equation}
Upon computing the eigendecomposition \eqref{eq:eigendecomposition}, the optimal strategy is given as
\begin{subequations}\label{eq:hyperbolic-strategy}
  \begin{equation}
    \tilde{\vartheta}_{\boldsymbol{y}}(s)  = \frac{1}{2} + \frac{1}{2}\mathrm{tanh}\left(\frac{s}{2}\right),
  \label{eq:hyperbolic-est}
  \end{equation}
  \begin{equation}
    \mathcal{M}_{\boldsymbol{y}}(s) = \mathcal{P}_{\boldsymbol{y}}(s),
  \label{eq:hyperbolic-pom}
  \end{equation}
\end{subequations}
where the optimal estimator is the logistic function.
Eqs.~\eqref{eq:hyperbolic-err} and \eqref{eq:hyperbolic-strategy} are the second result of this work---a quantum metrology for optimal weight estimation. 
Its application is next illustrated.

\emph{Estimation of a blend parameter.}---Consider 
the mixture 
\begin{equation}
  \rho_{\boldsymbol{\hat{y}}}(\eta) = \eta \ketbra{0} + (1-\eta)\tau_{\boldsymbol{\hat{y}}},
\label{eq:state-ex}
\end{equation}
where $\eta$ is the relative weight of the first component and $\tau_{\boldsymbol{\hat{y}}} = (\sigma_0 + \boldsymbol{\hat{y}}\cdot \boldsymbol{\sigma})/2$. 
Here, $\boldsymbol{\sigma} = (\sigma_1, \sigma_2, \sigma_3)$ are the Pauli matrices, $\sigma_3 \ket{0} = \ket{0}$, $\sigma_0$ is the identity matrix, and $\boldsymbol{\hat{y}}$ is a unit vector with azimuthal angle $ 0 \leq \alpha < 2\pi$ and polar angle $ 0 < \beta \leq \pi$.
We shall now address the optimal estimation of $\eta$. 

If only the hypothesis range $\theta \in (a, 1-a)$ is known \emph{a priori}, the initial state of information is captured by the normalised Haldane prior $p(\theta) = 1/[\kappa \theta(1-\theta)]$, with $\kappa = 4\mathrm{artanh}(1-2a)$.
Using this and Eq.~\eqref{eq:state-ex}, Eq.~\eqref{eq:hyperbolic-aux} renders the operators $\rho_{\boldsymbol{\hat{y}},0} = (\ketbra{0}+\tau_{\boldsymbol{\hat{y}}})/2$ and $\rho_{\boldsymbol{\hat{y}},1} = \chi(\ketbra{0}-\tau_{\boldsymbol{\hat{y}}})$, where
\begin{equation}
  \chi = -\frac{\log[a(1-a)]}{4} + \frac{\mathrm{Li}_2(a)-\mathrm{Li}_2(1-a)}{\kappa} 
\end{equation}
and $\mathrm{Li}_2(z)$ denotes the dilogarithm.
The optimisation equation \eqref{eq:lyaponuv} can then be solved by inspection upon noticing that $\tau_{\boldsymbol{\hat{y}}}^2 = \tau_{\boldsymbol{\hat{y}}}$. 
This leads to 
\begin{equation}
  \mathcal{S}_{\boldsymbol{\hat{y}}} = 2\chi(\ketbra{0}-\tau_{\boldsymbol{\hat{y}}}),
\label{eq:blend-solution}
\end{equation}
whose eigendecomposition reveals the optimal strategy 
\begin{subequations}\label{eq:blend-optimal}
  \begin{equation}
    \tilde{\vartheta}_{\boldsymbol{\hat{y}}}(s_{\pm})  = \frac{1}{2} + \frac{1}{2}\mathrm{tanh}\left(\frac{s_{\pm}}{2}\right),
    \label{eq:op-est_blend}
  \end{equation}  
  \begin{equation}
    \mathcal{M}_{\boldsymbol{\hat{y}}}(s) = \delta(s-s_{+}) \ketbra{s_{+}} + \delta(s-s_{-})\ketbra{s_{-}}.
  \end{equation}
\end{subequations}
Here, $s_{\pm} = \pm 2\chi\mathrm{sin}(\beta/2)$ and 
\begin{equation}
  \ket{s_\pm} = \frac{\mathrm{cos}(\beta/2)\ket{0} + [\mathrm{sin}(\beta/2)\mp 1]\mathrm{e}^{i \alpha}\ket{1}}{\sqrt{2[\mathrm{sin}(\beta/2)\mp 1]}}.
  \label{eq:op-pom_blend}
\end{equation}
Furthermore, the associated mean hyperbolic error is $\bar{\epsilon}_{\beta,\mathrm{min}} = \kappa^2/12 - 4\chi^2 \mathrm{sin}^2(\beta/2)$. 

To assess the relative precision gain, we compare $\bar{\epsilon}_{\beta,\mathrm{min}}$ to the prior error $\bar{\epsilon}_p = \kappa^2/12$ using $\varepsilon_{\beta} = \abs{\bar{\epsilon}_{\beta,\mathrm{min}}-\bar{\epsilon}_p}/\bar{\epsilon}_p = 48 \chi^2\mathrm{sin}^2(\beta/2)/\kappa^2$.
For fixed polar angle, the maximum gain is achieved in the limit of maximum ignorance: $\mathrm{lim}_{a\rightarrow 0}\,\varepsilon_{\beta} = 3\,\mathrm{sin}^2(\beta/2)/4 \leq 3/4$.
Eqs.~\eqref{eq:blend-optimal} can hence improve on a completely uninformed scenario as much as $75 \%$. 
On the other hand, precision gains become smaller as the local regime is approached, here realised when $a \sim 1/2$ and for which $\varepsilon_{\beta} \sim \mathrm{sin}^2(\beta/2)(2a-1)^2/3$.
This is because, the better the prior knowledge is, the harder it becomes for a measurement to improve on it. 
Note that precision gains become negligible when $\beta \ll 1$; indeed, Eq.~\eqref{eq:state-ex} would lose its dependency on $\eta$ should $\beta$ be allowed to vanish.

These precision gains can be exploited in practice by optimising individual shots in a finite sequence of them.
Imagine, for example, a protocol rendering the measurement outcomes $\boldsymbol{s} = (s_1,\dots,s_{\mu})$, where $s_i = s_{\pm}$.
Following Refs.~\cite{rubio2021global,rubio2023quantum}, the rule to simultaneously processing $\boldsymbol{s}$ into an optimal blend parameter estimate can be written as
\begin{equation}
  2 \tilde{\vartheta}_{\boldsymbol{\hat{y}}}(\boldsymbol{s}) = 1+\mathrm{tanh}\left[\int d\theta\,p(\theta|\boldsymbol{s},\boldsymbol{\hat{y}})\,\mathrm{artanh}(2\theta-1)\right],
\label{eq:repetitions}
\end{equation}
where $p(\theta|\boldsymbol{s},\boldsymbol{\hat{y}}) \propto p(\theta) \prod_{i=1}^{\mu} p(s_i|\theta,\boldsymbol{\hat{y}})$ is Bayes's theorem, and $p(s_i|\theta,\boldsymbol{\hat{y}}) = \bra{s_{\pm}}\rho_{\boldsymbol{\hat{y}}}(\theta)\ket{s_{\pm}}$.
This \emph{a priori} optimised approach has already been proven useful in Mach-Zehnder interferometry \cite{rubio2018quantum}, qubit sensing networks \cite{rubio2020bayesian}, and the aforementioned thermometry experiment on cold $^{41}$K atoms \cite{glatthard2022optimal}. 

\begin{figure}[t]
  \includegraphics[trim={0cm 0cm 0cm 0cm},clip,width=\linewidth]{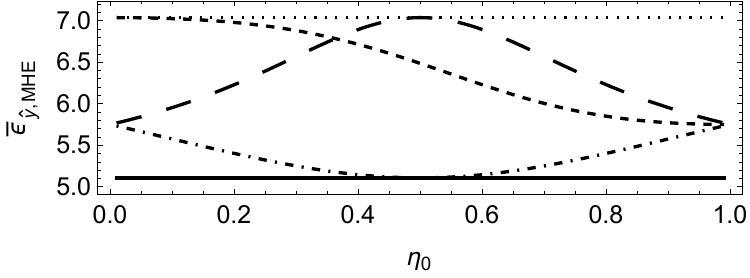}
  \caption{Mean hyperbolic errors for the estimation of $\eta$ in Eq.~\eqref{eq:state-ex} using \emph{local} measurements, i.e., given by the eigenstates of the symmetric logarithmic derivative $L_{\boldsymbol{\hat{y}}}(\eta_0)$ \cite{helstrom1976quantum, demkowicz2015quantum}. 
  Here, $\eta_0$ represents an initial guess at $\eta$, assumed to lie in the range $[0.01,0.99]$, and $\boldsymbol{\hat{y}}$ is a unit vector with azimuthal angle $\alpha$ and polar angle $\beta$.
  The latter is fixed as $\beta = \pi/2$. 
  Three azimuthal angles are chosen: $\alpha_1 = 0$ (dot-dashed line), $\alpha_2 = \pi/4$ (short-dashed line), and $\alpha_3 = \pi/2$ (long-dashed line).
  The prior error and the global minimum as per weight estimation correspond to the dotted and solid lines, respectively.
  Aside from the error for $\alpha_1$ at $\eta_0 = 1/2$, which saturates the minimum, every other configuration is suboptimal.
  Worse, no information is sometimes retrieved, as illustrated by the errors for $\alpha_2$ and $\alpha_3$ when $\eta_0 \rightarrow 0$ and $\eta_0 = 1/2$, respectively. 
  This contrasts with symmetry-informed estimation, which readily identifies Eq.~\eqref{eq:op-pom_blend} as the globally optimal POM.
  }
\label{fig:local_poms}
\end{figure}

It is instructive to further compare the performance of the optimal projectors \eqref{eq:op-pom_blend} with that of projecting onto the eigenspace of symmetric logarithmic derivatives (SLDs) $L_{\boldsymbol{\hat{y}}}(\eta_0)$, as is done in local estimation \cite{genoni2008optimal, demkowicz2015quantum}.
Here, $\eta_0$ is an initial `hint' at $\eta$ needed because the SLD is parameter dependent (the SLD is solution to $L_{\boldsymbol{\hat{y}}}(z)\rho_{\boldsymbol{\hat{y}}}(z) + \rho_{\boldsymbol{\hat{y}}}(z) L_{\boldsymbol{\hat{y}}}(z) = 2\partial_z \rho_{\boldsymbol{\hat{y}}}(z)$). 
Fig.~\ref{fig:local_poms} shows the numerical mean hyperbolic error, as a function of $\eta_0$, for the estimator \eqref{eq:op-est_blend} and three SLD POMs labelled by their azimuthal angle as $\alpha_1 = 0$ (dot-dashed line), $\alpha_2 = \pi/4$ (short-dashed line), and $\alpha_3 = \pi/2$ (long-dashed line).
For all of them, $\beta = \pi/2$ and $a = 0.01$.
The prior error ($\bar{\epsilon}_{p,f}$, dotted line) and the global minimum ($\bar{\epsilon}_{\beta, \mathrm{min}}$, solid line) are also shown. 
As can be seen, the POM $\alpha_1$ saturates $\bar{\epsilon}_{\beta, \mathrm{min}}$ at $\eta_0 = 1/2$, but it is increasingly less informative as $\eta_0 \rightarrow 0$, $1$. 
The POMs $\alpha_2$ and $\alpha_3$ are always suboptimal and uninformative for $\eta_0 \rightarrow 0$ and $\eta_0 = 1/2$, respectively, since the correspondent errors evaluate to $\bar{\epsilon}_{p,f}$.
Local estimation cannot thus always identify universally optimal measurements.

In summary, using symmetries in metrology can reduce the search for good strategies to finding the form of $f$ and performing a single calculation of the kind in Eqs.~\eqref{eq:blend-optimal}.
Metrological tasks such as identifying fundamental precision limits and informing the design of experimental protocols follow straightforwardly.
This is the final result.

\begin{table}[t]
  \setlength\extrarowheight{2pt}
  \begin{tabular}{lcc} 
  \hline\hline
  \multicolumn{3}{c}{\textbf{Minimal assumptions}} \\
  \hline
  \textbf{Parameter} & phase & location-isomorphic\\
  \textbf{Support} & $0  \leq \theta < 2\pi$ & $ -\infty < f(\theta) < \infty$ \\
  \hline
  \multicolumn{3}{c}{\textbf{Metrological formulation}} \\
  \hline
  \textbf{Invariance} & $\theta' = \theta + 2m\pi$ & $ f(\theta') = f(\theta) + c$ \\
  \textbf{Ignorance prior} & $p(\theta) = 1/2\pi$ & $p(\theta) \propto df(\theta)/d\theta$\\
  \textbf{Error} & $4\,\mathrm{sin}^2[(\tilde{\theta}-\theta)/2]$ & $[f(\tilde{\theta}) - f(\theta)]^2$ \\
  \hline\hline
\end{tabular}
\caption{
Symmetry-informed metrologies. 
Phase estimation applies to circular parameters (second column). 
For quantities isomorphic to location parameters, the prescription in the third column, together with Eqs.~\eqref{eq:lyaponuv}, \eqref{eq:optimal-strategy}, and \eqref{eq:min}, identifies the optimal strategies. 
It also unifies the metrologies of locations ($f(z) = z$), scales ($f(z) = \log(z/z_0)$, with constant $z_0$), and weights ($f(z) = 2\,\mathrm{artanh}(2z-1)$), and it gives the theoretical support needed to discover new metrologies under minimal assumptions.
Note that $m \in \mathbb{Z}$ and $c \in \mathbb{R}$. 
}
\label{tab:metrologies}
\end{table}

\emph{Concluding remarks.}---Symmetry-informed estimation is put forward as a universally optimal framework for location-isomorphic metrology.
Eqs.~\eqref{eq:lyaponuv}, \eqref{eq:optimal-strategy}, and \eqref{eq:min} enable the direct calculation of the best estimator and POM, together with the corresponding minimum error.
Having made minimal assumptions, these apply to any parameter range, prior information, or state, including multiple copies \cite{demkowicz2020multiparameter}. 
Furthermore, fixed-POM estimators such as Eq.~\eqref{eq:repetitions} indicate that the notion of location-isomorphic parameter is also relevant for classical measurements. 
Despite its single-shot formulation, this framework is straightforward to use in practice, either by repeating an \emph{a priory} optimised strategy, as in Eq.~\eqref{eq:repetitions}, or using adaptive schemes \cite{brivio2010experimental, mehboudi2021fundamental} where each shot is optimised by maximising the precision gain \eqref{eq:gain}.
In general, this will reduce the number of runs needed to achieve a good precision in experiments measuring location-isomorphic parameters, thus enabling a better resource allocation.

Combining this framework with phase estimation [Tab.~\eqref{tab:metrologies}] offers an unprecedented extension of the class of exactly solvable problems in Bayesian metrology. 
This covers ubiquitous quantities such as phases, locations, scales, and weights, but also any other parameter type for which invariance of our initial state of knowledge under Eq.~\eqref{eq:basic-transformation} holds.
For instance, correlation coefficients ranging from $-1$ to $1$, or if invariance under reparametrisations of some statistical model is desired (using information geometry, this leads to $f(z) = \int^z dt \mathcal{F}(t)^{1/2}$, where $\mathcal{F}$ is the Fisher information \cite{jorgensen2021bayesian}).
Moreover, this capacity to accommodate physical symmetries enables the rigorous application of quantum metrology to fundamental problems such as the detection of dark matter \cite{bassi2022way,sherrill2023analysis}.
Overall, symmetry-informed estimation is not unlike the use of symmetries to derive the correct Euler-Lagrange equations in theoretical mechanics.

% Acknowledgements
\emph{Acknowledgments}. The author gratefully thanks S. Bukbech, F. Albarelli, W. G\'{o}recki, D. Branford, N. Sherrill, J. Boeyens, S. Michaels, L. A. Correa, and M. Perarnau-Llobet for helpful comments, and the participants of the QUMINOS workshop for insightful discussions.  
Parts of this manuscript were written during a visit to the Open Quantum Systems Group at the University of La Laguna. 
This work was funded by the Surrey Future Fellowship Programme.

% References
\bibliography{refs}

\end{document}